\documentstyle[12pt]{article}
\begin{document}

\begin{flushright}
{\bf BIHEP-TH-94-10}\\
{\bf  April, 1994}
\end{flushright}

\begin{center}
 {\large\bf	Light Quark Dependence of the Isgur-Wise  Function   from
 QCD Sum rules\footnote{This project was supported by the National Science
Foundation of China and the Grant LWTZ-1298 of the Chinese Academy of
Science.}}
\end{center}
\vspace{0.5cm}
\begin{center}
{\bf Tao  Huang$^{a,b}$  and  Chuan-Wang Luo}$^b$\\
a.{\small\sl CCAST (World Laboratory), P.O.Box 8730, Beijing 100080, China }\\
b.{\small\sl Institute of High Energy Physics, P.O.Box 918(4), Beijing 100039,
China\footnote{Mailing address}}
\end{center}
\vspace{0.6cm}

\begin{abstract}
	We study light quark dependence of the Isgur-Wise function for
$B_a\rightarrow D_a$ and $B_a\rightarrow D^{*}_a$ in the framework of
QCD sum rules. At zero recoil , all the Isgur-Wise functions equal as
required by heavy quark symmetry and at non-zero  recoil,  the Isgur-Wise
function for $B_s$ decay falls faster than that for
 $B_{u,d}$ decay , which is just contrary to the recent prediction of the
heavy meson chiral perturbation theory. As  by-products,  we also estimate
SU(3) breaking effects in the mass and the decay constant.

\end {abstract}

\vspace{1.8cm}

{\it PACS number(s):  11.50.Li,12.38.Cy,12.38.Lg,13.25.+m }
\newpage

\small
\begin{flushleft}
{\Large\bf I $~$ Introduction} \\
\end{flushleft}

	Recently, there has been a great deal of interest in weak decays
of heavy mesons made from one heavy quark and one light quark.	As heavy quark
goes into infinite mass limit, all form factors for $B\rightarrow D$ and
$B\rightarrow D^{*}$ can be expressed in terms of a single universal function
[1], the so-called Isgur-Wise function [2]. It is the property of the
Isgur-Wise function--the normalization at the zero-recoil point [1,2,3] that
leads to the model independently
extraction of the Kabayashi-Makawa matrix element $V_{cb}$. The knowledge of
the Isgur-Wise function will be required  for any phenomenological applications
of heavy quark symmetry to exclusive weak decays of heavy mesons. Therefore,
it is very interesting and
necessary to investigate the Isgur-Wise function and its properties.

	The Isgur-Wise function represents the nonperturbative dynamics of weak
 decays of heavy mesons. It depends not only on the dimensionless product
 $v\cdot v^{\prime}$ of the initial and final mesonic velocities, but also on
 the light quark flavor of the initial and final mesons [4,5].  In the past
few years,
 many different nonperturbative methods were developed to investigate the
velocity
 product $v\cdot v^{\prime}$ dependence of the Isgur-Wise function [6-13]. But
for light quark flavor dependence of the Isgur-Wise function, Only chiral
perturbation theory of the heavy mesons [14] was developed to study it [5],
no  investigation  from other methods can be found in the literatures .
However, QCD sum rule approach [15] involving nonperturbative effects
is based on the QCD theory of strong interaction,  it has many advantages
over others:  it is not only  suitable for investigations of $v\cdot
v^{\prime}$
 dependences of the Isgur- Wise function but also suitable for investigating
 its light quark flavor dependence.
 In [10-13], QCD sum-rule was used to calculate the $v\cdot v^{\prime}
 $ dependence of the Isgur-Wise  function.  In this paper, we will use QCD sum
 rules to calculate its light quark flavor dependence.

	This paper is organized as follows: section II presents the sum rule
 for the decay constant. The sum rule for the Isgur-Wise function is derived
in section III.  Section IV gives  the numerical analysis and the final section
is reserved  for  Summary and discussion.

 \begin{flushleft}
 {\large\bf II.$~$ Sum rule for the decay constant}
 \end {flushleft}

	In order to investigate light quark flavor dependence of the Isgur-Wise
 function ,  one should know the light quark flavor dependence of the decay
 constant of the heavy meson. i.e., to derive the sum rule for decay constant
 including the light quark dependence in the heavy quark effective theory
(HQET).

       In HQET, the low energy parameter $F_a(\mu)$ of heavy meson
$M_a(\bar{q}Q)$ is defined by [10,16]
\begin{equation}
 <0|\bar{q}\Gamma h_{Q}|M_a(v)>\;=\;\frac{F_a(\mu)}{2} Tr[\Gamma M(v)],
 \end{equation}
where  $M(v)$ is the spin wavefunction of heavy meson $M_a(v)$ in HQET
\begin{equation}
M(v)\;=\;\sqrt{m_{Q}}\frac{1+\not v}{2} (-i\gamma_{5}).
\end{equation}
In leading order ,the decay constant $f_{M_a}\simeq F_a(\mu)/\sqrt{m_{M_a}}$.
It should be emphasized that here and after, the subscript $ a=u,d,s$ specifies
the light antiquark flavor $\bar{q}=\bar{u},\bar{d},\bar{s}$ of the heavy meson
$M_a(\bar{q}Q)$.

      The standard procedure to calculate the physical quantity with QCD sum
rule can be found in [15]. In this paper, we   follow the method in [10] given
by Neubert.
To derive the sum rule for the decay constant , one can consider the
two-point correlation function in HQET,
\begin{equation}
    \pi_5(\omega)\;=\;i\int d^4 x e^{i k\cdot x}<0|T{A^{(v)}_5(x),
A^{(v)+}_5(0)}|0>,
\end{equation}
where $A_5^{(v)}=\bar{q}\gamma_{5} h_{Q}$ is the effective peseudoscalar
current and $\omega=2 k\cdot v$  , k is the residual momentum .The starting
point of QCD sum rule is to calculate the correlation function
$\pi_{5}(\omega)$
in two different ways. First, in not so deep Euclidean region of $\omega$,
where
nonperturbative effects enter but do not dominate, by the operator product
expansion(OPE), one can expand $\pi_{5}(\omega)$ as
\begin{equation}
\pi_{5}(\omega)\;=\;\pi^{P}_{5}(\omega)+\pi^{NP}_{5}(\omega).
\end{equation}
The first term $\pi_{5}^{P}(\omega)$ is just the usual perturbative
contribution, which corresponds to the identity operator in the OPE , and
is expressed through a dispersion relation
\begin{equation}
\pi_{5}^{P}(\omega)\;=\;\frac{1}{\pi}\int ds \frac{\rho_P (s)}{s-\omega}
+ subtractions.
\end{equation}
The perturbative spectral density $\rho_{p}(s)$ can be computed as usual
\begin{equation}
\rho_{P}(\omega)\;=\;\frac{3}{8\pi}\sqrt{\omega^2-4 m_{q}^2}(\omega+
2 m_{q})\Theta(\omega-2 m_{q})
\end{equation}
( $m_{q}$ is the mass of the light quark q).
The second term $\pi_{5}^{NP}(\omega)$ is the nonperturbative
contributions, according to SVZ [15], which are parameterized
by the quark condensate, the gluon condensate and the quark-gluon
mixed condensate etc, and we obtain
\begin{eqnarray}
\pi_{5}^{NP}(\omega) & = & <0|\bar{q}q|0> [\frac{1}{\omega}+\frac{m_{q}}
{2\omega^2}+\frac{m_{q}^2}{\omega^3}]
  \nonumber  \\
& &+<0|\frac{\alpha_s}{\pi}GG|0> \frac{m_q}{2\omega^3}[1-ln\frac{\omega}{\mu}]
  \nonumber \\
 & & -\frac{g_{s}}{2\omega^3}<0|\bar{q}\sigma Gq|0>
   \nonumber \\
 & & + \frac{8\pi\alpha_s}{27\omega^4}<0|\bar{q}q|0>^2.
\end{eqnarray}

	On the other hand, the correlation function can also be reexpressed
in terms
of hadronic resonance states and continuum states by a dispersion relation as
\begin{equation}
\pi_5^{Ph}(\omega)\;=\;\frac{1}{\pi}\int^{\infty}_{\omega^c_a} ds \frac{
\rho_H (s)}{s-\omega-i\epsilon} + \frac{|<0|\bar{q}\gamma_5 h_Q(v)|M(v)>|^2}
{(2 \bar{\Lambda}_a-\omega-i\epsilon)m_Q}+subtractions,
\end{equation}
where $\bar{\Lambda}_a = M_a-m_{Q}$ is the parameter of HQET [17] and
$\omega^c_{a}$ is the threshold of the continuum
states. Assuming the quark-hadron duality, the continuum spectral function
$\rho_{H}$ can be approximated by the perturbative spectral function
$\rho_{p}$.
Therefore  one gets the sum rule
\begin{equation}
\pi_5^{Th}(\omega)\;=\;\pi_5^{Ph}(\omega) + subtractions.
\end{equation}
In order to enhance contribution of the lowest lying resonance state and
improve the convergence of the OPE , the Borel transformation defined as
\begin{equation}
\frac{1}{T} \hat{B}_T^{(\omega)}\;=\; \lim_{\scriptstyle -\omega,~n\rightarrow
\infty \atop  \scriptstyle  T=-\omega/n~ fixed}
   \frac{\omega^n}{\Gamma (n)}[-\frac{d}{d\omega}]^n
\end{equation}
must be applied . So  the final sum rule reads
\begin{eqnarray}
  F_a^2(\mu) e^{-2\bar{\Lambda}/T} & =
&\frac{3}{8\pi^2}\int^{\omega^c_a}_{2m_q}
ds \sqrt{s^2-4m^2_q}[2 m_q + s]e^{-s/T}
 \nonumber  \\
& &  -<0|\bar{q}q|0>[1-\frac{m_q}{2T}+\frac{m^2_q}{2T^2}]
-\frac{<0|\frac{\alpha_s}{\pi}GG|0>m_q}{4T^2}[\gamma-0.5-ln\frac{T}{\mu}]
   \nonumber   \\
& & +\frac{g_s <0|\bar{q}\sigma Gq|0>}{4T^2}+ \frac{4\pi\alpha_s}{81T^3}
<0|\bar{q}q|0>^2  \\
&=& G( T^{-1}),
\end{eqnarray}
with $\gamma=0.5772 $ being the Euler constant. Taking the derivative with
respect to the inverse of T ,  one can  obtain the sum rule for
$\bar{\Lambda}_a$
\begin{equation}
\bar{\Lambda}_a\;=\;-\frac{G^{\prime}(T^{-1})}{2G(T^{-1})}.
\end{equation}

	From these sum rules , one can observe that the light quark flavor
dependence of $F_{a}(\mu)$ is represented by that of the condensates
$<0|\bar{q}q|0>$ , $<0|\bar{q}\sigma Gq|0>$ and $m_{q}$. In the next section ,
these sum rules will be used to obtain the sum rule for the Isgur-Wise
function.

\begin{flushleft}
{\large \bf III.Sum rule for the Isgur-Wise function}
\end{flushleft}

	The Isgur-Wise function $\xi_a(v\cdot v^{\prime},\mu)$ is defined
 by the matrix element at the leading order in $\frac{1}{m_Q}$[10,18]
\begin{equation}
<M_a(v^{\prime})|\bar{h}_{Q_2}(v^{\prime})\Gamma h_{Q_1}(v)|M_a(v)>\;=\;
-\xi_a(v\cdot v^{\prime}, \mu) Tr[\bar{M}(v^{\prime})\Gamma M(v)],
\end{equation}
which is valid for  an arbitrary matrix $\Gamma$ .

	To derive the sum rule for the Isgur-Wise function , one should
consider the three-point correlation function in HQET
\begin{equation}
\tilde{\pi}(\omega,\omega^{\prime},y)\;=\; \int d^4x d^4z e^{i(k^{\prime}
\cdot x-k\cdot z)} <0|T \{[\bar{q}\gamma_5 h_{Q_1}(v^{\prime})]_x,
[\bar{h}_{Q_1}(v^{\prime})\Gamma h_{Q_2}(v)]_0,
[\bar{h}_{Q_2}(v)\gamma_5 q]_z \}
\end{equation}
with $y=v\cdot v^{\prime}$ ,$\omega=2 k\cdot v $ and$\omega^{\prime}=2
k^{\prime}\cdot v^{\prime} $.  The weak current in Eq.(15) is
$\bar{h}_{Q_1}(v^{\prime})\Gamma h_{Q_2}(v)$.
	To be convenient, let's factorize out the Lorentz structure by
defining $\tilde{\pi}(\omega,\omega^{\prime},y)=\pi(\omega,\omega^{\prime},y)
Tr[\frac{1+\not v}{2}\Gamma\frac{1+\not v^{\prime}}{2}]$.
the perturbative spectral density  $\rho_{pert}$ for $\pi^P(\omega,\omega^{
\prime},y)$ is

\begin{equation}
\rho_{pert}(\omega,\omega^{\prime},y)=\frac{3}{16\pi}\frac{[\omega+
\omega^{\prime}+2(1+y)m_q]}{(1+y)\sqrt{y^2-1}}\theta(\omega)
\theta(\omega^{\prime}) \theta [2y\omega\omega^{\prime}-\omega^{\prime 2}
-\omega^2-4 m^2_q(y^2-1)]  \nonumber    \\
\end{equation}

	The next standard step is to write the correlation function by using
dispersion relations in $\omega$ and $\omega^{\prime}$
\begin{eqnarray}
\pi^{Ph}(\omega,\omega^{\prime},y)\;=\;&\frac{\xi_a(y,\mu) F^2_a(\mu)}{(
2\bar{\Lambda}_a-\omega-i\epsilon)(2\bar{\Lambda}_a-\omega_{\prime}-i\epsilon)}
+\frac{1}{\pi} \int^{\infty}_{\omega^c}ds \int^{\infty}_{\omega^c}ds^{\prime}
\frac{\rho_H(s,s^{\prime},y)}{(s-\omega-i\epsilon)(s^{\prime}-\omega^{\prime}
-i\epsilon)} \nonumber \\
 & +subtractions
\end{eqnarray}
with $\rho_H(s,s^{\prime},y)=\rho_{pert}(s,s^{\prime},y)$ by assuming the
quark-hadron duality.

	After applying Borel transformations with respect to $\omega$ as
well as $\omega^{\prime}$ to improve the matching between $\pi^{Ph}(\omega,
\omega^{\prime},y)$ and $\pi^{Th}(\omega,\omega^{\prime},y)$
\begin{equation}
\hat{B}^{(\omega)}_{\tau^{\prime}}\hat{B}^{(\omega)}_{\tau}
\pi^{Ph}(\omega,\omega^{\prime},y)\;=\;
\hat{B}^{(\omega)}_{\tau^{\prime}}\hat{B}^{(\omega)}_{\tau}
\pi^{Th}(\omega,\omega^{\prime},y),
\end{equation}
we obtain the sum rule
\begin{equation}
\xi_a(y,\mu) F^2_a(\mu)e^{-2\bar{\Lambda_a}/T}\;=\;\frac{1}{\pi}
\int^{\omega^c}_{0}ds
\int^{\omega^c}_{0}ds^{\prime}\rho_{Pert}(s,s^{\prime},y)
e^{-(s+s^{\prime})/2T}
+\pi_B^{NP}(y,T),
\end{equation}
where  we have set $\tau^{\prime}=\tau=2T$ as observed in [19]. The borelized
nonperturbative contribution $\pi^{NP}_{B}(y,T)$ is
\begin{eqnarray}
\pi^{NP}_{B}(y,T) &=& \hat{B}^{(\omega)}_{\tau^{\prime}}\hat{B}^{(\omega)}_{
\tau}\pi^{NP}(\omega,\omega^{\prime},y)|_{\tau^{\prime}=\tau=2T}
\nonumber  \\
&=& -<0|\bar{q}q|0>[1-\frac{m_q}{2T}+\frac{m_q^2}{4T^2}(1+y)]
 \nonumber   \\
& & + <0|\frac{\alpha_s}{\pi}GG|0>[\frac{y-1}{48T(1+y)}-\frac{m_q}{4T^2}
(\gamma-0.5-\ln\frac{T}{\mu})]    \nonumber  \\
& & + \frac{g_s<0|\bar{q}\sigma Gq|0>}{4T^2}\frac{2y+1}{3}+
\frac{4\pi\alpha_s <0|\bar{q}q|0>^2}{81T^3} y.
\end{eqnarray}

	Since the integration domain is symmetric in s and $s^{\prime}$
, changing variables $\alpha=\frac{s+s^{\prime}}{2} $, $\beta=s-s^{\prime}$
and improving the continuum threshold model as suggested in [10] by Neubert,
we get the sum rule for the Isgur-Wise function:
\begin{equation}
\xi_a(y,\mu)\;=\;\frac{K(T,\omega^c_a,y)}{K(T,\omega^c_a,1)},
\end{equation}
where
\begin{eqnarray}
K(T,\omega^c_a,y) = & \frac{3}{8\pi^2}(\frac{2}{1+y})^2  \int^{\omega^c}_{m_q
\sqrt{2(1+y)}}   d\alpha [\alpha+(1+y)m_q]\sqrt{\alpha^2-2(1+y)m^2_q}
 e^{-\alpha/T}   \nonumber   \\
 &  -<0|\bar{q}q|0>[1-\frac{m_q}{2T}+\frac{m^2_q}{4T^2}(1+y)]  \nonumber  \\
 & + <0|\frac{\alpha_s}{\pi}GG|0>[\frac{y-1}{48T(1+y)}-\frac{m_q}{4T^2}
(\gamma-0.5-\ln\frac{T}{\mu})]  \nonumber   \\
 & + \frac{g_s<0|\bar{q}\sigma Gq|0>}{4T^2}\frac{2y+1}{3}+
\frac{4\pi\alpha_s <0|\bar{q}q|0>^2}{81T^3} y.
\end{eqnarray}
In the above derivation,  we have used the sum rule  for $F_a(\mu)$.

\begin{flushleft}
{\large \bf IV.Numerical analysis}
\end{flushleft}

	In the numerical analysis of sum rules,	we take the following values
for parameters such as condensates and $m_q$[20-24]
\begin{eqnarray}
& <0|\bar{u}u|0>=<0|\bar{d}d|0>=(-0.23GeV)^3 &
   \nonumber   \\
&<0|\bar{u}\sigma G u|0>=<0|\bar{d}\sigma G d|0>=0.8GeV^2 <0|\bar{u}u|0> &
       \nonumber  \\
& \frac{<0|\bar{s}s|0>}{<0|\bar{u}u|0>}=\frac{<0|\bar{s}\sigma G s|0>}
{<0|\bar{u}\sigma G u|0>}=0.8~~~~~ ;~~~~~ <0|\frac{\alpha_s}{\pi}GG|0>
=0.012 GeV^4 &   \nonumber   \\
& m_u\approx m_d\approx 0~~~~~ ;~~~~~ m_s\simeq 0.15 GeV &
\end{eqnarray}
and set the scale $\mu=1GeV$, which equals about two times of
$\bar{\Lambda}_{u,d,s}$ (see below). For the continuum model $\omega^c=
\sigma (y)\omega^c_a$, we use the experiment preferred model $\sigma(y)=
\frac{y+1}{2y}$ as suggested in [10] by Neubert.

	As $a=u,d$, all sum rules for $\bar{\Lambda}_a$, $F_a$ and $\xi_a$
have been evaluated in [10,16,18]. In the following , we will evaluate these
sum rules as $a=s$ and calculate the ratios $R_F=F_s/F_{u,d}$ and
$R_{IW}=\xi_s/ \xi_{u,d}$.

	In Fig.1, we show $\bar{\Lambda}_s$ and $F_s$ as a function of T for
different $\omega^c_s $. Within $\omega^c_s=1.8\sim 2.4 GeV $ and $T=0.6\sim
1.0 GeV$ , where QCD sum rules calculation is reliable, we have
\begin{equation}
\bar{\Lambda}_s\simeq 0.62\pm 0.07 GeV~~~,~~~ F_s\simeq 0.36\pm 0.05 GeV^{3/2}.
\end{equation}

For completeness, we  list the values for $\bar{\Lambda}_{u,d}$ and
$F_{u,d}$:
\begin{equation}
\bar{\Lambda}_{u,d}\simeq 0.55\pm 0.07 GeV~~~ ,~~~F_{u,d}\simeq 0.32
\pm 0.05 GeV^{3/2}.
\end{equation}

Therefore, with the above values $\bar{\Lambda}_a$ and $F_a$, one can calculate
the $SU(3)$ breaking effects in the mass of the heavy meson $M(\bar{q}_a Q)$ to
the leading order in $1/m_{Q}$
\begin{equation}
\Delta M\;=\;m_{M_s}-m_{M_{u,d}}\;=\;\bar{\Lambda}_s-\bar{\Lambda}_{u,d}
\end{equation}
and the ratio $R_F=F_s/F_{u,d}$. However, in order to reduce the errors,
writing the mass difference $\Delta M=\bar{\Lambda}_s-\bar{\Lambda}_{u,d}$
and the ratio $R_F=F_s/F_{u,d}$ with the corresponding sum rules,
in Fig.2, one can find that $\Delta M$ and $R_F$ depend on T very weakly
within $\omega_{u,d}^c=1.7\sim2.3GeV$ and $\omega_s^c=1.8\sim 2.4 GeV$.
{}From Fig.2(a) follows
\begin{equation}
\Delta M=69\pm 5 MeV,
\end{equation}
which is in good agreement with the recent experiment results [25,26]
\begin{equation}
m_{B_s}-m_{B}=90 \pm 6 MeV, \nonumber \\
m_{D_s}-m_{D}=99.5 \pm 0.6 MeV.
\end{equation}

{}From Fig.2(b), we get the ratio
\begin{equation}
R_F=1.13 \pm 0.01.
\end{equation}

	In Fig.3, the Isgur-Wise function $\xi_s$ is shown as a function of y.
Changing  $\omega_s^c$ in $1.8\sim 2.4 GeV$ and T in
$0.7\sim 0.9GeV$, the Isgur-Wise function varys in the band region. Obviously,
the dependence on these parameters is very weak.
At the center of the sum rule window  T=0.8GeV,  we obtain the slope
parameter $\varrho^2_a$ defined as
$\varrho^2_a=-\xi^{\prime}_a(y=1,\mu)$
\begin{equation}
\varrho^2_s\;=\;1.09\pm 0.04,
\end{equation}
the uncertainty is due to the variation of$\omega^c_s$.
One can compare with
\begin{equation}
\varrho^2_{u,d}\;=\;1.01\pm 0.02.
\end{equation}
and find that SU(3) breaking effects in the slope parameter is not
 large but the important thing is
\begin{equation}
\varrho^2_s > \varrho^2_{u,d}.
\end{equation}
This result just indicates that the Isgur-Wise function $\xi_s$ falls faster
than the Isgur-Wise function $\xi_{u,d}$ as shown below. Obviously, all these
slope parameters satisfy the Bjorken sum rule $\varrho^2_a > 0.25$ [27] but
violate the Voloshin sum rule $\varrho^2_a < 0.75$\footnote{However, it
should be emphassed that this violation of the Voloshin sum rule depends on the
 choice of $\sigma (y)$. If choosing $\sigma (y)=1$ as discussed in
[12], one finds $\varrho^2_a $ can satisfy the Voloshin sum rule.}[28].

	In Fig.4, we show $R_{IW}=\xi_s/\xi_{u,d}$ as a function of y at
$T=0.8 GeV$ for different $\omega^c_{u,d}=1.7\sim 2.3 GeV$ and $\omega_s^c
=\omega^c_{u,d}+0.1GeV$. One can find that the ratio $R_{IW}$ displays a soft
dependence on $\omega^c_{u,d,s}$ .

	In order to show how $R_{IW}$ depends on T, in Fig.5,  we plots
$R_{IW}$ as a function of T at $ y=1.6$, which approximately corresponds to
the largest recoil point $q^2=0$ for $B_{u,d}\rightarrow D_{u,d}+l\nu $.
In the stable region, we get
\begin{equation}
R_{IW}\simeq (95\pm 2 )\%,
\end{equation}
where the uncertainty is ascribed to the uncertainty in $\omega^c_{u,d,s}$
and T.

	It should be pointed out that in the evaluations of sum rules
for $\xi_a$ and $R_{IW}$, the continuum model is chosen as $\sigma (y)
=\frac{y+1}{2y}$. This may cause large errors in $ \xi_{a}$ and $ R_{IW}$.
As discussed in [10], one knows
\begin{equation}
\frac{y+1-\sqrt{y^2-1}}{2}\leq \sigma (y) \leq 1,
\end{equation}
and the model $\sigma_{max}=1$ and $\sigma_{min}=\frac{y+1-\sqrt{y^2-1}}{2}$
respectively constituents the upper bound and the lower bound for $\xi_a$ .
For $R_{IW}$, as shown in Fig.6, the model $\sigma_{max}$ and $\sigma_{min}$
just gives the lower bound and the upper bound respectively.  Although
different continuum model gives different value for $R_{IW}$, one can find
that all of these values clearly give
\begin{equation}
R_{IW} < 1~~,~~ for~ y\not = 1.
\end{equation}
Therefore we conclude that $R_{IW} < 1 $(for $y\not = 1$) is independent
of the model choice  $\sigma (y)$.

\begin{flushleft}
{\large \bf V. Summary and Discussion.}
\end{flushleft}

	In summary , we have determined the parameters $\bar{\Lambda}_s$ and
 $F_s$, and given the Isgur-Wise function $\xi_s(y,\mu)$.
Also we have shown how large SU(3) breaking effects exist in the mass,the decay
constant and the Isgur-Wise function. It is very interesting to find that
the Isgur-Wise function for $ B_s\rightarrow D_s$ falls faster than the
Isgur-Wise function for $B_{u,d}\rightarrow D_{u,d}$, which is just contrary
to the prediction of the heavy meson chiral perturbation theory where only
SU(3) breaking chiral loops are calculated [5]. Our result $R_{IW}\leq 1$
agrees with that of the BSW model [6]. It is expected that the future
experiments can test this result and reveal the underlying mechanism of
 SU(3) breaking effects.

\begin{center}
{\large \bf  Acknowledgment}
\end{center}

	One of us (C.W.Luo) would like to thank M.Neubert for helpful
discussion.

\newpage

\begin{flushleft}
{\large \bf  Figure Captions}
\end{flushleft}

Fig.1: $\bar{\Lambda}_s$ and $F_s$ as a function of T for different
$\omega^c_s$:
       Dashed line: $\omega_s^c=1.8GeV$, Solid line: $\omega^c_s=2.1GeV$,
       Dotted line: $\omega^c_s=2.4GeV$.   \\

Fig.2: The mass difference $\Delta M=\bar{\Lambda}_s-\bar{\Lambda}_{u,d}$ and
       the ratio $R_F=F_s/F_{u,d}$ as a function of T
($\omega^c_s=\omega^c_{u,d}
	+0.1GeV$): Dashed line: $\omega_{u,d}^c=1.7GeV$, Solid line:
	$\omega^c_{u,d}=2.0GeV$, Dotted line: $\omega^c_{u,d}=2.3GeV$.\\

Fig.3: The Isgur-Wise function $\xi_s$ as a function of y . The band
corresponds
       to variations of 	$\omega_s^c$ in $ 1.8GeV\sim 2.4 GeV$ and T in $
       0.7GeV\sim 0.9GeV $.\\

Fig.4: The ratio $R_{IW}=\xi_s/ \xi_{u,d}$ as a function of y at T=0.8GeV
	($\omega^c_s=\omega^c_{u,d}+0.1GeV$): Dashed line:
	$\omega_{u,d}^c=1.7GeV$, Solid line:
	$\omega^c_{u,d}=2.0GeV$, Dotted line: $\omega^c_{u,d}=2.3GeV$. \\

Fig.5: The ratio $R_{IW}$ as a function of T at
y=1.6($\omega^c_s=\omega^c_{u,d}
	+0.1GeV$): Dashed line: $\omega_{u,d}^c=1.7GeV$, Solid line:
	$\omega^c_{u,d}=2.0GeV$, Dotted line: $\omega^c_{u,d}=2.3GeV$.   \\

Fig.6:  The ratio $R_{IW}=\xi_s/ \xi_{u,d}$ as a function of y at T=0.8GeV
	for different models $\sigma_{max}$ (Fig.6(a)) and $\sigma_{min}$
	(Fig.6(b)). ($\omega^c_s=\omega^c_{u,d}+0.1GeV$): Dashed line:
	$\omega_{u,d}^c=1.7GeV$, Solid line:
	$\omega^c_{u,d}=2.0GeV$, Dotted line: $\omega^c_{u,d}=2.3GeV$. \\

\end{document}